\documentclass[12pt]{article}
\usepackage{epsf}
\usepackage{amsmath}
\usepackage{graphics}
\usepackage{cite}

\renewcommand{\thefootnote}{\#\arabic{footnote}}
\begin{document}

\newcommand{\gtrsim}{ \mathop{}_{\textstyle \sim}^{\textstyle >} }
\newcommand{\lesssim}{ \mathop{}_{\textstyle \sim}^{\textstyle <} }

\newcommand{\rem}[1]{{\bf #1}}

\renewcommand{\thefootnote}{\fnsymbol{footnote}}
\setcounter{footnote}{0}
\begin{titlepage}

\def\thefootnote{\fnsymbol{footnote}}

\begin{center}
\hfill November 2014\\
\vskip .5in
\bigskip
\bigskip
{\Large \bf Bang or Bounce}

\vskip .45in

{\bf Paul H. Frampton}

{\em Courtyard Hotel, Whalley Avenue, New Haven, CT 06511, USA.}

\end{center}

\vskip .4in
\begin{abstract}
Following up an earlier suggestion of how the Tolman Entropy
Conundrum (TEC) can be solved in a cyclic cosmology
using the Come Back Empty (CBE) assumption with phantom
dark energy, here we show
how the same CBE strategy may work with a cosmological
constant in the expansion era.
As in the earlier case, this leads to a multiverse,
actually an infiniverse, with the concomitant issues of
predictivity and testability.
Here we show how extreme flatness and homogeneity
at the bounce are natural properties of the contraction
era, interestingly without any necessity for an inflationary
era at the beginning of the present expansion.
Essential ingredients in the solution of TEC are CBE contraction
and a careful treatment of what is meant by the visible
universe.
\end{abstract}
\end{titlepage}

\renewcommand{\thepage}{\arabic{page}}
\setcounter{page}{1}
\renewcommand{\thefootnote}{\#\arabic{footnote}}

\newpage

\noindent
{\it Introduction.} It is of broad interest to understand better the nature of
the early universe especially the big bang. The discovery\cite{Penzias}
of the cosmic microwave background(CMB) in 1965 resolved a dichotomy
then existing in theoretical cosmology between steady-state and big 
bang theories. The interpretation of the CMB as a relic of a big bang was
compelling and the steady-state theory died. Actually at that time it was
really a trichotomy being reduced to a dichotomy because a third theory,
a bounce in a cyclic cosmology had been under study since 1922
\cite{Friedmann}. 

\bigskip

\noindent
Nevertheless, for purely theoretical reasons, the bounce 
had been discarded due to the Tolman Entropy Conundrum(TEC)
\cite{Tolman1,Tolman2}. The TEC, stated simply, is that the entropy
of the universe necessarily increases, due to the second law of
thermodynamics, and therefore cycles become larger and longer in the
future, smaller and shorter in the past, implying that a big bang must
have occurred at a finite time in the past.

\bigskip

\noindent
Some progress towards a solution of the TEC was made in \cite{BF}
using the Come Back Empty (CBE) assumption in the BF model.
A huge entropy was there jettisoned at turnaround and the significantly
smaller universe, empty of matter, contracted adiabatically to a bounce
with zero entropy. The BF model employed so-called phantom dark energy
with equation of state $\omega < -1$. Since this violates energy conditions,
it is more conservative to use a cosmological constant with $\omega = -1$
as we do here. We shall find CBE is still possible.

\bigskip

\noindent
Assuming the cosmological principle of homogeneity and isotropy leads
to the FLRW metric
\begin{equation}
ds^2 = dt^2 - a(t)^2 \left[ \frac{dr^2}{1-kr^2} + r^2 (d\theta^2 + \sin^2\theta d\phi^2) \right]
\end{equation}
where $a(t)$ is the scale factor and $k$ the curvature. Inserting this metric
into the Einstein equation leads to two Friedmann equations. The first is the expansion
equation
\begin{equation}
H(t)^2 = \left(\frac{\dot{a}(t)}{a(t)} \right)^2 = \frac{8\pi G}{3} \rho_{TOT}(t) - \frac{k}{a(t)^2}
\label{Friedmann1}
\end{equation}
where $\rho_{TOT}(t)$ is the total density.

\bigskip

\noindent
Using the continuity equation $\dot{\rho} + 3H(\rho+p)=0$, differentiation
gives rise to a second equation
\begin{equation}
\frac{\ddot{a}(t)}{a(t)} = - \frac{4\pi G}{3} (\rho + 3p)
\label{Friedmann2}
\end{equation}

\bigskip

\noindent
The critical density is defined by $\rho_c(t) = [3H(t)^2/8\pi G]$ and discussion
of flatness involves proximity to one of the quantity
\begin{equation}
\Omega_{TOT} (t) = \frac{\rho_{TOT}}{\rho_c(t)}
\end{equation}

\bigskip

\noindent
Let the present time be $t=t_0=1.38\times10^{10}$y. Normalize $a(t_0)=1$ and define
$H(t_0)=H_0$. Other relevant times all measured relative to the
would-have-been bang at $t=0$ are the Planck time $t_{Planck}=10^{-44}$s,
the electroweak time $t_{EW}=10^{-10}$s, the onset of matter domination
$t_m=4.7\times10^4y$, the onset of dark energy domination $t_{DE}=9.8$Gy
and for the turnaround time we shall use $t_T=t_0+150=163.8$Gy. The bounce occurs
at $t_B$ where $t_{Planck}<t_B<t_{EW}$.

\bigskip

\noindent
In the unadorned big bang theory, one has dependences of the scale factor $a(t) \sim t^{1/2}$
for $t_{Planck}<t<t_m$, $a(t)\sim t^{2/3}$ for $4t_m<t<t_{DE}$, and
$a(t) = \exp[H_0(t-t_0)]$ for $t_{DE}<t<t_T$. The leads to the values of $a(t)$
to be used later:$a(t_{Planck})=2.3\times10^{-32}$, $a(t_{EW})=2.3\times10^{-15}$,
$a(t_m)=2.8\times10^{-4}$, $a(t_{DE}) =0.75$, $a(t_0)=1$, and $a(t_T)=5.7\times10^4$.

\bigskip

\noindent
Two of the most striking observations of the universe are isotropy to an
accuracy $1\pm O(10^{-5})$ and flatness $\Omega_{TOT}(t_0)=1.00\pm0.05$.
In big bang theory this implies that in the early universe $\Omega_{TOT}(t)$ is given by
\begin{equation}
\Omega_{TOT}(t_{EW}) = 1 \pm O(10^{-27})
\label{flat1}
\end{equation}
\begin{equation}
\Omega_{TOT}(t_{Planck}) = 1 \pm O(10^{-61})
\label{flat2}
\end{equation}

\bigskip
\bigskip

\noindent
{\it Inflation.}  A simple way to present inflation is to rearrange Eq.(\ref{Friedmann1})
after division by $H(t)^2$ as
\begin{equation}
(\Omega_{TOT}(t) - 1) = \frac{k}{\dot{a}(t)^2}
\label{Flatness}
\end{equation}
In a decelerating expansion the denominator of the RHS in
Eq.(\ref{Flatness}) becomes small and $\Omega_{TOT}(t)$ deviates
more and more from $\Omega_{TOT}(t)=1$. This is why the proximity
of $[\Omega_{TOT}(t_0)-1]$ to zero imposes the strong initial conditions in
Eqs.(\ref{flat1}) and (\ref{flat2}). 

\bigskip

\noindent
The most popular explanation of flatness is inflation\cite{Guth1}
which inserts a period of highly accelerated superluminal expansion
at time $t=t_{inflation}$ during the period
$t_{Planck}<t_{inflation}<t_{EW}$. While inflating, $\dot{a}(t)$ becomes extremely
large enforcing flatness so precisely that the subsequent decelerating
expansion during $t_{inflation}<t<t_{DE}$ does not remove it.

\bigskip

\noindent 
Inflation also explains homogeneity and has other successes
including the prediction of 
scale-invariant density perturbations and of the reddening
spectral index. Inflation has been incorporated into string theory
\cite{KKLT,KKLMMT}. The successful discovery of the BEH boson adds
credibility to the existence of the scalar inflaton. One possible objection to
inflation, that it leads to eternal inflation\cite{Guth2} and hence to
a multiverse in which predictivity is hampered by the measure
problem\cite{Vilenkin}, is not a fatal flaw.

\bigskip

\noindent
Thus, only a compelling alternative could cast doubt on the 
correctness of inflation.

\bigskip
\bigskip

\noindent
{\it Turnaround.}
To solve the TEC, the entropy of the visible universe must
essentially vanish as discussed in \cite{BF}. As given above, the scale factor
at turnaround is $a(t_T)=5.7\times10^4$ at $t_T=163.8$Gy. At the present time 
the visible universe has radius $R_{VU}=4.4\times10^{26}$m. For $t=t_T$, this evolves to
$R_{VU}(t_T)\simeq R_{VU}(\infty)=R_{VU}(t_0)+cH_0^{-1}=5.7\times10^{26}$m
due to the exponential expansion. 

\bigskip

\noindent
Because of the superluminal expansion
of space $R_{VU}(t_0)$ is meanwhile stretched to the much larger radius
$a(t_T)R_{VU}(t_0) = 2.5\times10^{31}$m. Assuming $10^{12}$ galaxies inside
$R_{VU}(t_0)$ a typical intergalactic(IG) distance is
$d_{IG}(t_0)=10^{-4}R_{VU}(t_0)=2.2\times10^{22}$m
which will be stretched to $d_{IG}(t_T)=a(t_T)d_{IG}(t_0) =2.5\times10^{27}$m
that is greater than $R_{VU}(t_T)$. This implies that a visible universe at turnaround
contains zero or one galaxies.

\bigskip

\noindent
It now requires great care to identify the appropriate visible universe at
turnaround to fulfill the CBE assumption. One might be tempted to include
the galaxy we live in but that would be too Ptolemaic, The correct choice of
$R_{VU}(t_T)$ is instead a sphere which contains no matter, luminous or dark, and no
black holes. It contains instead only dark energy with no entropy and an 
infinitesimal quantity of both curvature and radiation, this last necessary
for the ensuing derivation of flatness.

\bigskip

\noindent
In the language of \cite{BF}, an important ratio ($f$) of the contraction to expansion
scale factors in $\hat{a}(t) = f a(t)$ is here given by
\begin{equation}
f=\frac{R_{VU}(t_T)}{a(t_T)R_{VU}(t_0)} = 2.3\times 10^{-5}
\label{f}
\end{equation}

\bigskip
\bigskip

\noindent
{\it Bounce.} The CBE contracting universe contains no matter.
It does, however, contain a crucial infinitesimal amount of radiation so that
approaching the bounce $\Omega_{TOT}(t)=\Omega_{\gamma}(t)$. During the expansion era
at $t=t_0$ its contribution is $\Omega_{\gamma}(t_0) = 1.3\times 10^{-4}$.

\bigskip

\noindent
The contraction Friedmann equation is, from Eq.(\ref{Friedmann2}) and using 
the radiation equation of state
$3p=\rho$
\begin{equation}
\frac{\ddot{\hat{a}}(t)}{\hat{a}(t)} = - \frac{8\pi G}{3} \rho_{\gamma}(t)
\label{contraction}
\end{equation}
and we need to calculate $\dot{\hat{a}}(t_B)$ which appears in this contraction
version of Eq.(\ref{Flatness})
\begin{equation}
\left| \Omega_{TOT}(t_B) - 1 \right| = \left| \frac{k}{\dot{\hat{a}}(t)_B)^2} \right|
\end{equation}

\bigskip

\noindent
Integrating Eq.(\ref{contraction}) and using $\rho_{\gamma}(t) = \rho_{\gamma 0}/\hat{a}(t)^4$
gives the following results for flatness at $t=t_{EW}$ and $t_{Planck}$
\begin{equation}
\left| \Omega_{TOT}(t_B) - 1 \right| = 3.5\times 10^{-46} ~~~ (t_B = t_{EW})
\end{equation}
\begin{equation}
\left| \Omega_{TOT}(t_B) - 1 \right| = 3.5\times 10^{-80} ~~~ (t_B = t_{Planck})
\end{equation}
which is our main result. 

\bigskip

\noindent
This extreme flatness is not surprising because it is suppressed
by the CBE factor $f^4 \sim 10^{-18}$ relative to the expanding phase together
with the stability of $\Omega_{TOT}(t)=1$ under contraction, behaving precisely oppositely
to the fine tunings of Eqs(\ref{flat1}) and (\ref{flat2}). Homogeneity at the bounce is
ensured by its essentially zero entropy.

\bigskip

\noindent
Thus, this provides an alternative explanation for the observed isotropy
and flatness which can now be regarded as evidence not for inflation but
for a bounce.

\bigskip
\bigskip

\noindent
{\it Discussion.} In order to construct a cyclic cosmology, a first requirement is
to address the TEC which was an impossible-seeming hurdle and which discouraged
workers in the 1920s who were otherwise very enthusiastic about this alternative
to the big bang. The only logical possibility is that a huge entropy must be jettisoned at
turnaround.

\bigskip

\noindent
It is interesting that TEC+CBE necessitates an accelerated expansion
just as was first discovered in 1998. With decelerated expansion the fraction $f<1$
necessary for the CBE assumption is not possible. The advantages of a bounce over
a bang are that the low entropy at the bounce ensures homogeneity and
that the appropriate degree of flatness is obtained naturally, interestingly
without the necessity of an inflation at the beginning of the expansion era.
It is almost inevitable that inflation will lead to eternal inflation
as explained in \cite{Guth2} and to a multiverse with its measure problem\cite{Vilenkin}.
Similarly this TEC+CBE model, just as in \cite{BF}, will almost inevitably lead to an
infiniverse; after all, the entropy at turnaround must go elsewhere.

\bigskip

\noindent
One prediction is that, because of the additional factor $f^4$, the flatness condition
at the present time is $\Omega(t_0)=1$, accurate to some eighteen decimal places.
Predictions for density perturbations and the spectral index are to be investigated
in future research.

\bigskip

\noindent
Finally it is worth remarking that in his monumental work\cite{Tolman1,Tolman2}, 
Tolman did not consider accelerated
expansion and, to my knowledge, never entertained the possibility of more
than one universe.

\end{document}